\documentclass[12pt]{article}
\usepackage{amscd,amsmath,amssymb,amstext,amsthm,exscale,latexsym}
\usepackage{graphicx}
\newcommand {\al}   {\alpha}       
\newcommand {\g }   {\gamma}

\newcommand {\pl}   {\partial}     

\newcommand   {\const}{{\sf\,const}}     \newcommand   {\diag}{{\sf\,diag\,}}
\newcommand   {\ex}{{\sf\,e}}            
\newcommand {\MC}  {{\mathbb C}}   
\newcommand {\ME}  {{\mathbb E}}   
   \newcommand {\MH}  {{\mathbb H}}
\newcommand {\MM}  {{\mathbb M}}   
   \newcommand {\MP}  {{\mathbb P}}
\newcommand {\MR}  {{\mathbb R}}   

\newcommand {\MU}  {{\mathbb U}}

\newcommand {\CL}  {{\cal L}}      
\newcommand {\Sa}  {{\textsc{a}}}   \newcommand {\Sb}  {{\textsc{b}}}

\begin{document}
\title     {On geometric interpretation of the Aharonov--Bohm effect}
\author    {M. O. Katanaev
            \thanks{E-mail: katanaev@mi.ras.ru}\\ \\
            \sl Steklov Mathematical Institute,\\
            \sl ul.~Gubkina, 8, Moscow, 119991, Russia}
\date      {\today}
\maketitle
\begin{abstract}
A geometric interpretation of the Aharonov--Bohm effect is given in terms of
connections on principal fiber bundles. It is demonstrated that the principal
fiber bundle can be trivial while the connection and its holonomy group are
nontrivial. Therefore, the main role is played by geometric rather than
topological effects.
\end{abstract}
\section{Introduction}
The Aharonov--Bohm effect \cite{AhaBoh59} has attracted great attention of
theorists and experimenters for many years. The interest is caused by the
following circumstance. In the theory of gauge fields, the popular belief is
that only nontrivial field strength rather than the potentials themselves that
are not gauge-invariant can cause the observable effects. Contrary to this
opinion, Aharonov and Bohm have demonstrated that the integral of a gauge field
along a closed loop can produce the observable effects. The effect produced by
the magnetic potential was soon confirmed experimentally
\cite{WerBri60,Chambe60,BoHaWoGr60}.

In the present work, we give a geometric interpretation of the Aharonov--Bohm
effect in terms of connections on the principal fiber bundle. Observable
effects are produced by holonomy group elements for the corresponding
connections. The elements of the holonomy group are gauge invariant because the
group $\MU(1)$ is Abelian one. Many authors relate the Aharonov--Bohm effect to
the nontrivial topology of space. We here demonstrate that the principal fiber
bundle can be trivial, while the connection arising on it generally has a
nontrivial holonomy group and hence leads to the observable effects. From this
it follows that the Aharonov--Bohm effect has geometric rather than topological
nature.

The Aharonov--Bohm effect from the geometric viewpoint is similar to the Berry
phase \cite{Berry84} (see also \cite{Katana11C}). In both cases, the observable
effects are produced by holonomy group elements. The difference is that bases of
the corresponding principal fiber bundles are different manifolds. For the
Aharonov--Bohm effect, the base is the four-dimensional space-time while for the
Berry phase, it is a manifold of external parameters (for example, the range of
variation of the external magnetic field) on which the Hamiltonian depends.
\section{Aharonov--Bohm effect}
The Aharonov--Bohm effect \cite{AhaBoh59} gives an example of occurrence of a
nontrivial connection on the trivial principal fiber bundle
$\MP\big(\MR^4,\pi,\MU(1)\big)\approx\MR^4\times\MU(1)$ in nonrelativistic
quantum mechanics. In this case, unlike the Berry phase, the space-time $\MR^4$
in which the particle moves acts as the base $\MM$ of the principal fiber
bundle.

Let us consider the Schr\"odinger equation \cite{Schrod26A,Schrod26B}
\begin{equation}                                                  \label{escheq}
  i\hbar\frac{\pl\psi}{\pl t}=H\psi,
\end{equation}
where $\hbar$ is the Planck constant. Let the Hamiltonian describe the motion of
a point particle of mass $m$ in the three dimensional Euclidean space $\MR^3$
with the Cartesian coordinates $x^\mu$, $\mu=1,2,3$,
\begin{equation}                                                  \label{ezehal}
  H_0=-\frac{\eta^{\mu\nu}p_\mu p_\nu}{2m}+U=-\frac{\hbar^2}{2m}\triangle+U,
\end{equation}
where $p_\mu=i\hbar\pl_\mu$ is the particle momentum operator,
$\eta_{\mu\nu}=\diag(---)$ is the negative definite space metric,
$\triangle=\pl_1^2+\pl_2^2+\pl_3^2$ is the Laplace operator, and $U(x)$ is the
potential energy of the particle.

We write down the four dimensional momentum operator in the form
$p_\al=i\hbar\pl_\al$, $\al=0,1,2,3$. In this case, the zero 4-momentum
component $p_0=i\hbar\pl_0=i\hbar\pl_t$ has the physical meaning of the
particle energy operator.

If a particle interacts with an electromagnetic field, this interaction is
described by means of the minimal substitution for all four components of the
momentum
\begin{equation}                                                  \label{emisus}
  p_\al~~\mapsto~~i\hbar\pl_\al-\frac ecA_\al,
\end{equation}
where $e$ is the particle charge, $c$ is the velocity of light, and $A_\al$ is
the electromagnetic field potential (components of the local form of the
$\MU(1)$-connection). In this case, the zero component divided by the velocity
of light, $A_0/c$, has the physical meaning of the electrical field potential,
and the spatial components $A_\mu$ are the covector components of the magnetic
field potential. Thus, the point particle moving in the electromagnetic field is
described by the Schr\"odinger equation
\begin{equation}                                                  \label{eshrch}
  i\hbar\frac{\pl\psi}{\pl t}=\left[\frac{\hbar^2}{2m}
  \eta^{\mu\nu}\left(\pl_\mu+i\frac e{\hbar c}A_\mu\right)
  \left(\pl_\nu+i\frac e{\hbar c}A_\nu\right)+\frac ecA_0\right]\psi+U\psi.
\end{equation}
For simplicity, we further put $\hbar=1$ and $c=1$.

From the geometric viewpoint, minimal substitution (\ref{emisus}) coincides to
within constants with replacement of the partial derivative by the covariant
one:
\begin{equation*}
  \pl_\al~~\mapsto~~\pl_\al+ieA_\al.
\end{equation*}

Let us consider Schr\"odinger equation (\ref{eshrch}) from the geometric
viewpoint. It is solved in the whole space-time $\psi=\psi(t,x)$; therefore, the
base of the fiber bundle (for example, see \cite{KobNom6369}) is the
four-dimensional Euclidean space $(t,x)\in\MR\times\MR^3=\MR^4$. In this case,
$\MR^4$ is considered simply as a four-dimensional manifold without any
four-dimensional metric. If required, the metric can be introduced but its
presence does not alter the principal fiber bundle and the connection. The
metric $\eta_{\mu\nu}$ is defined only on spacelike sections $t=\const$, because
it enters the Schr\"odinger equation. The wave function $\psi(t,x)$ is the
section of the fiber bundle $\ME\big(\MR^4,\pi_\ME,\MC,\MU(1),\MP\big)$ with the
typical fiber being the complex plane $\MC$, which is associated with a certain
principal fiber bundle $\MP\big(\MR^4,\pi,\MU(1)\big)$. This principal fiber
bundle is always trivial $\MP\approx\MR^4\times\MU(1)$, because its base is the
four-dimensional Euclidean space. The local $\MU(1)$-connection form defined by
the electromagnetic potential $A_\al(t,x)$, is set on this principal fiber
bundle. In nonrelativistic quantum mechanics, instead of the whole set of
sections of associated fiber bundles, we consider only the subset $\psi(t,x)$
consisting of such differentiable sections which belong to the Hilbert space of
square integrable functions $\MH=\CL_2(\MR^3)$ on spacelike sections  $\MR^3$
for each moment of time $t$.

We consider two cases.
\subsection{Electric potential}
Let us assume that the magnetic potential is equal to zero, $A_\mu=0$,
$\mu=1,2,3$. We now write down the Schr\"odinger equation in the form
\begin{equation}                                                  \label{eschre}
  i\dot\psi=(H_0+eA_0)\psi,
\end{equation}
where $H_0$ is the Hamiltonian of the system without the electromagnetic
potential (\ref{ezehal}), and the dot atop designates differentiation with
respect to time. We also assume that the electric potential depends only on
time, $A_0=A_0(t)$. We seek a solution of Schr\"odinger equation (\ref{eschre})
in the form $\psi=\ex^{-i\Theta}\phi$ where $\phi$ is the solution of the free
Schr\"odinger equation:
\begin{equation}                                                  \label{esheqf}
  i\dot\phi=H_0\phi,
\end{equation}
and $\Theta=\Theta(t)$ is a certain phase independent of the space point.
Substitution of $\psi=\ex^{-i\Theta}\phi$ into initial Schr\"odinger equation
(\ref{eschre}) leads to the equation for the phase
\begin{equation*}
  \dot\Theta=eA_0,
\end{equation*}
where we have canceled the common phase multiplier $\ex^{-i\Theta}$ and $\phi$.
This can be done, because the Schr\"odinger equation must be fulfilled for all
$t$ and $x$. The solution of this equation has the form
\begin{equation}                                                  \label{ewavab}
  \Theta(t)=\Theta_0+e\int_0^t\!\!\!dsA_0(s),
\end{equation}
where $\Theta_0$ is the value of the wave function phase at the initial moment
of time.

\begin{figure}[h,b,t]
\hfill\includegraphics[width=.70\textwidth]{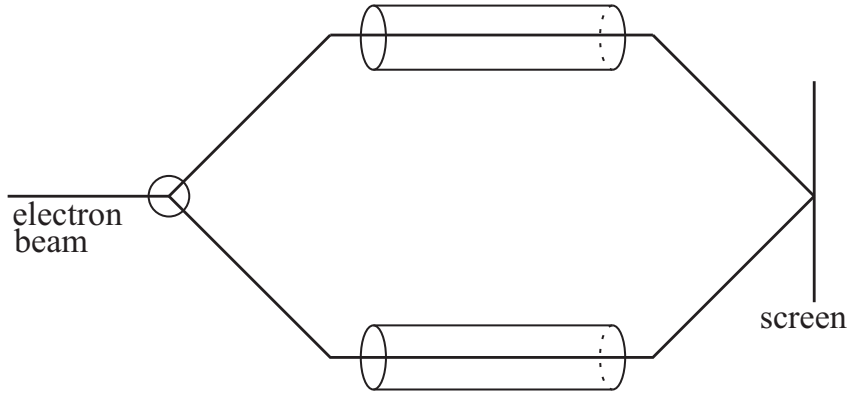}
\hfill {}
\\
\centering \caption{Scheme of the Aharonov--Bohm experiment.\label{figure1}}
\end{figure}

Aharonov and Bohm proposed the experiment whose scheme is shown in
Fig.\ref{figure1}. The beam of electrons is split into two beams transmitted
through two metal tubes to which different potentials are applied. Then beams
are collected, and the interference pattern is observed on the screen.
The electric potential applied to the tubes depends on time. It is supposed
to be zero until both beams are inside the tubes. Then it increases to some
values different for each tube, and then decreases to zero when beams leave the
tubes. Thus, the beams are in the field $A_0$ only when they are inside the
tubes.

The interference pattern depends on the electron phase difference in the beams,
which can be estimated approximately as follows. Assume that an electron is
described by the wave function $\psi(t,x)$ which satisfies Schr\"odinger
equation (\ref{eschre}) in the whole space-time $\MR^4$. We consider that at
each moment of time, the wave function support is nonzero in a small
neighborhood of space near the particle trajectory. Only in this case it is
possible to speak about the particle trajectory. In particular, when the
electron passes through the metal tube, it is assumed that the support of the
wave function lies completely in the tube. Mathematically, this can be described
by the corresponding potential introduced into Schr\"odinger equation
(\ref{eschre}). This hypothetical potential does not change the space-time,
that is, the base of the principal fiber bundle; it only provides the electron
motion along the present trajectory. We now estimate the electron phase change
in the upper beam. Since the electric field potential is homogeneous inside the
tube and the support of the wave function is completely localized inside the
tube, we can consider that the phase of the wave function is defined by integral
(\ref{ewavab}). We designate the moments of time the beam splits and achieves
the screen by $t_1$ and $t_2$, respectively. Then the phase of the electron
wave function of the upper beam, when it approaches the screen, changes by the
value determined by the integral:
\begin{equation*}
  \Theta_1=e\int_{t_1}^{t_2}\!\!\!dtA^{(1)}_0(t),
\end{equation*}
where $A_0^{(1)}(t)$ is the electric field potential at the moment of time $t$,
that is, at that point of space where the electron of the upper beam is located
at the moment of time $t$. Analogously, the phase of the wave function of an
electron of the lower beam changes by
\begin{equation*}
  \Theta_2=e\int_{t_1}^{t_2}\!\!\!dtA^{(2)}_0(t),
\end{equation*}
where $A_0^{(2)}$ is the electric field potential along the lower trajectory. It
is clear that the phase difference for the electrons of the upper and lower
beams $\Theta_{\Sa\Sb}=\Theta_2-\Theta_1$ can be written in the form of the
integral
\begin{equation}                                                  \label{eahboe}
  \Theta_{\Sa\Sb}=e\oint_\g dtA_0(t).
\end{equation}
along the closed contour $\g$ in the space-time, when the lower half of the
contour shown in Fig.\ref{figure1} is passed first and then the upper half is
passed in the opposite direction. Figure \ref{figure1} shows the projection
of the contour $\g$ onto the space plane.

Let us return to the geometric interpretation. The electron phase difference is
given by integral (\ref{eahboe}), which is unambiguously defined by the contour
$\g$ and the potential $A_0$ set on it. The electric potential $A_0$ is the time
component of the local $\MU(1)$-connection form on the principal fiber bundle
$\MP\big(\MR^4,\pi,\MU(1)\big)$. Therefore, phase difference (\ref{eahboe})
defines the element of the holonomy group
$\ex^{i\Theta_{\Sa\Sb}}\in\Phi\big((t_1,x_1),e\big)\subset\MU(1)$ at the point
$(t_1,x_1)$, where $x_1$ is the space point where the beam is split and
$e$ is the unity of the group. The base of the principal fiber bundle $\MP$ has
trivial topology. Therefore, the contour $\g$ can be always contracted to a
point, metallic tubes being at their location, and the contour continuously
passing through them. For contracting contour, the Aharonov--Bohm phase
$\Theta_{\Sa\Sb}$ goes to zero, and the respective element of the holonomy group
goes to unity $e\in\MU(1)$. This is well known, because the holonomy group is
the Lie subgroup in the structure group. We note that contraction of the contour
is not related to the experiment where the definite element of the holonomy
group corresponding to the initial location of the contour is observed.

The proposed geometric interpretation is not unique. Since the support of the
wave function is supposed to be located in a small neighborhood of the
trajectory, many principal fiber bundles $\MP'\big(\MM,\pi,\MU(1)\big)$ with
different bases $\MM$ can be constructed. To this end, it is sufficient to take
the fiber bundle $\MP\big(\MR^4,\pi,\MU(1)\big)$ with a given connection and to
narrow down the base by removing any arbitrary number of arbitrary located
straight lines in space $\MR^3$ which are perpendicular to the plane of
Fig.\ref{figure1} and lie outside of the contour $\g$. As a result, the base
$\MM$ becomes not simply connected manifold. The given procedure appears to be
unnatural. Moreover, it results in a much more complicated problem since initial
Schr\"odinger equation (\ref{eshrch}) in that case must be solved on a
topologically nontrivial manifold.

At the end of this section, we consider the transformation of the local
$\MU(1)$-connection form under changing of the section. From Schr\"odinger
equation (\ref{eschre}) it follows that under the vertical automorphism
\begin{equation*}
  \psi'=\ex^{ia}\psi,
\end{equation*}
where $a=a(t)$ is a differentiable function of time, components of the local
$\MU(1)$-connection form are transformed according to the rule
\begin{equation*}
  eA'_0=eA_0+\dot a,
\end{equation*}
as components of the local $\MU(1)$-connection form.

Thus we see that the Aharonov--Bohm effect and the Berry phase are based on the
nontrivial geometry, i.e., on the connection with nontrivial holonomy group
rather than on the topology. In this case, the space topology can be both
trivial and nontrivial.
\subsection{Magnetic potential}
Now we consider the case when the electric field potential is equal to zero,
$A_0=0$. Suppose that the magnetic field potential depends only on space
coordinates $x^\mu$ and is independent on time $t$ (a static field). Then
the Schr\"odinger equation takes the form
\begin{equation}                                                  \label{eshmaf}
\begin{split}
  i\dot\psi&=\frac1{2m}\eta^{\mu\nu}(\pl_\mu+ieA_\mu)
  (\pl_\nu+ieA_\nu)\psi+U\psi
\\
  &=\frac1{2m}\eta^{\mu\nu}(\pl^2_{\mu\nu}\psi+2ieA_\mu\pl_\nu\psi
  +ie\pl_\mu A_\nu\psi-e^2A_\mu A_\nu\psi)+U\psi.
\end{split}
\end{equation}
Let $\phi$ be a solution of the Schr\"odinger equation in the absence of
electromagnetic field (\ref{esheqf}). Then it is not difficult to verify that
the function
\begin{equation*}
  \psi=\ex^{-i\Theta}\phi,
\end{equation*}
where the phase $\Theta$ satisfies the equation
\begin{equation}                                                  \label{emagpa}
  \pl_\mu\Theta=eA_\mu,
\end{equation}
is a solution of initial Schr\"odinger equation (\ref{eshmaf}).

\begin{figure}[h,b,t]
\hfill\includegraphics[width=.70\textwidth]{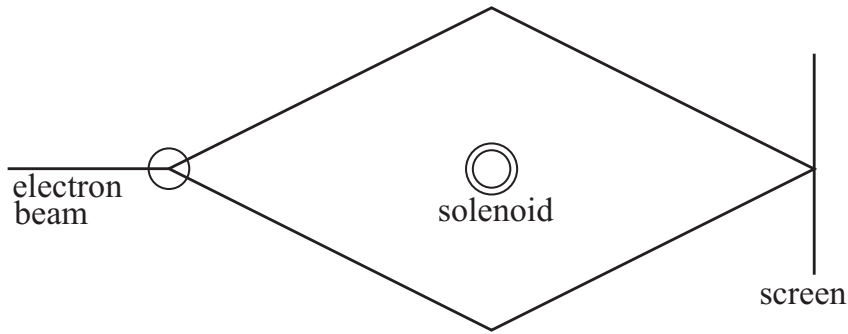}
\hfill {}
\\
\centering \caption{Scheme of the Aharonov--Bohm experiment for the
determination of the phase $\Theta$.\label{figure2}}
\end{figure}

Aharonov and Bohm proposed the experiment for the determination of the phase
$\Theta$ whose scheme is shown in Fig.\ref{figure2}. In this experiment, a
beam of electrons is split into two beams bending from different sides an
infinitely long solenoid with constant magnetic flux $\Phi$, which is
perpendicular to the plane of the figure. Then beams are collected, and the
interference pattern depending on the phase difference of electrons in different
beams is observed.

To estimate the phase difference of electrons, we make the same assumptions as
in the case of the electric field; namely, we consider that the Schr\"odinger
equation without magnetic potential has a solution with the support which is
concentrated in a small neighborhood of the electron trajectory. We suppose that
this property can be provided by introduction of the appropriate potential in
Eq.(\ref{esheqf}). This potential does not change the topology of space-time,
but only ensures motion of electrons along the given trajectory. Then
Eqs.(\ref{emagpa}) are fulfilled for the phase of the solution of the
Schr\"odinger equation with magnetic potential. Since the magnetic field
vanishes outside the solenoid, $\pl_\mu A_\nu-\pl_\nu A_\mu=0$, the
integrability conditions for system of equations (\ref{emagpa}) are fulfilled.
Therefore, the phase difference can be expressed through the contour integral
\begin{equation}                                                  \label{eahbkm}
  \Theta_{\Sa\Sb}=e\oint_\g dx^\mu A_\mu,
\end{equation}
where $\g$ is a closed contour in four-dimensional space-time surrounding the
solenoid. We note that the term $dx^0A_0$ in the integrand is equal to zero
because $A_0=0$ by assumption. This integral is independent of the chosen
contour surrounding the solenoid, because the magnetic field outside of the
solenoid vanishes.

The Aharonov--Bohm phase, using the Stokes formulae, can be written in the form
of the surface integral
\begin{equation}                                                  \label{eahbom}
  \Theta_{\Sa\Sb}=\frac12e\iint_Sdx^\mu\wedge dx^\nu F_{\mu\nu}=e\Phi,
\end{equation}
where $F_{\mu\nu}$ is the magnetic field strength (components of the local
curvature 2-form) and $\Phi$ is the total magnetic flux through the solenoid.
We note that to use the Stokes formulae, we must consider that the magnetic
field is defined everywhere in space $\MR^3$, including the solenoid itself.

The geometric interpretation of the Aharonov--Bohm effect consists in the
following. We have the same principal fiber bundle as in the case of electric
potential, $\MP\big(\MR^4,\pi,\MU(1)\big)$, whose base is the four-dimensional
Euclidean space $(t,x)\in\MR\times\MR^3=\MR^4$, in which electrons move and the
structure group is the unitary group $\MU(1)$ (the phase factor $\ex^{i\Theta}$
of the wave function). However, the connection is now different: only space
components of the local connection form $A_\mu$, $\mu=1,2,3$, differ from zero.
Aharonov--Bohm phase difference (\ref{eahbkm}) is uniquely defined by the
contour $\g$ (the same as for the electric potential) and values of the
connection form $A_\mu$ on it. Writing the contour integral in the form of
surface integral (\ref{eahbom}), we assume that the connection is defined on the
whole space-time $\MR^4$. This means that we consider solenoid of finite radius
to avoid singularities.

Thus, the principal fiber bundle is trivial, and the Aharonov--Bohm phase
$\Theta_{\Sa\Sb}$, depending on the connection and the contour, defines uniquely
the element of the holonomy group. When the contour is contracted to a point,
the corresponding element of the holonomy group goes to unity $e\in\MU(1)$ as it
must be. Here we assume that the contour goes freely through the solenoid.

As for the electric potential, the electron wave function is the cross section
of the associated fiber bundle $\ME\big(\MR^4,\pi_\ME,\MC,\MU(1),\MP\big)$ with
the Euclidean space $\MR^4$ as the base, and the structure group is the unitary
group $\MU(1)$. Under the vertical automorphism,
\begin{equation*}
  \psi'=\ex^{ia}\psi,
\end{equation*}
where $a=a(x)$ is a differentiable function of spatial coordinates $x^\mu$,
$\mu=1,2,3$, and the magnetic field potential is transformed according to the
rule
\begin{equation*}
  eA'_\mu=eA_\mu-\pl_\mu a.
\end{equation*}
This follows from Schr\"odinger equation (\ref{eshmaf}). Thus, components
of the magnetic field potential have indeed the behavior of components of the
local connection form.

Since electron phase difference (\ref{eahbkm}) is defined by values of the local
connection form only near the integration contour, the base of the trivial
principal fiber bundle $\MP\big(\MR^4,\pi,\MU(1)\big)$ can be narrowed down
without modification of the answer. For example, we can cut out the domain
of the space-time which lies inside the contour $\g$ and contains the solenoid.
Then the base in no longer simply connected. For this reason the Aharonov--Bohm
effect is often called topological. As already demonstrated above, this is not
necessary. It is sufficient to assume that the magnetic field differs from zero
only on a finite domain on the plain in Fig.\ref{figure2} inside the integration
contour. If we assume that the base of the fiber bundle is the Euclidean space
$\MR^4$ with the solenoid cut out, this will create additional difficulties,
because the Schr\"odinger equation must be solved on not simply connected
manifold. Moreover, the Stokes formulae cannot be used in such case. Thus, the
Aharonov--Bohm effect induced by the magnetic potential must be considered as
geometric rather then topological.

The Aharonov--Bohm effect with electric and magnetic potentials attracts great
attention of physicists for the following reason. According to contemporary
points of view, the gauge-invariant functions are the only observables in gauge
models. From this point of view, the electromagnetic field potential $A_\al$,
$\al=0,1,2,3$, is not observable itself, because it is not gauge-invariant. In
the considered cases, the electron beams are not subjected to the action of
electromagnetic forces, since the electric and magnetic field strengths are zero
in the regions where electrons move. Therefore, it seems that the difference in
phases of electron beams has to be zero. However, from the Schr\"odinger
equation it follows that it is not so. It should be noted that the
electromagnetic field potential is not observable, but we observe its integral
along the closed contour which defines the element of the holonomy group of
the $\MU(1)$-connection being a gauge-invariant object, because the gauge group
is Abelian one.
\section{Conclusions}
In this paper, we have given the geometric interpretation of the Aharonov--Bohm
effect. It was demonstrated that the space-time topology in this case can be
trivial. Therefore, the Aharonov--Bohm effect, like the Berry phase, has the
geometric rather then topological nature.

The interpretation proposed in the paper contains nothing except standard
differential geometric notions. In a geometric interpretation of mathematical
physics models, one has to take into account that a connection exists on any
principal fiber bundle independently of the base topology \cite{KobNom6369}.
Moreover, if a family of local connection forms is given on an arbitrary closed
submanifold of the base of some principal fiber bundle, then the corresponding
connection can always be continued to the whole principal fiber bundle. This can
be done in many ways. The connection defines the holonomy group which generally
is nontrivial.

In experiments for measuring the Aharonov--Bohm effect, the observed effects
are produced not by the whole holonomy group, but a fixed element of the
holonomy group which depends on the connection and the closed contour. The
base topology may be trivial or not, it does not play any role. If the topology
is trivial, the integration contour can be contracted to a point. The effect
disappears in this case because the corresponding element of the holonomy group
goes to the unity element, and this is quite natural from the physical
viewpoint.

A connection on the principal fiber bundle defines connections on all fiber
bundles which are associated with it. In particular, if the typical fiber is an
infinitely dimensional Hilbert space, then the connection is also defined. At
present, the interpretation of the Berry phase and the Aharonov--Bohm effect, as
a rule, is reduced to consideration of a connection on an associated fiber
bundle, and this forces one to consider infinitely dimensional manifolds and to
take into account the related subtleties. From our point of view, the
interpretation of geometric effects in terms of connections on principal fiber
bundles is simpler and more natural.

The author is grateful to I.~V.~Volovich for discussions and fruitful comments.

This work was supported in part by the Russian Foundation for Basic Research
(grants 11-01-00828-a and 11-01-12114-ofi\_m), the Program for Supporting
Leading Scientific Schools (Grant No.\ NSh-7675.2010.1), and the program
``Contemporary problems of theoretical mathematics'' of the Russian Academy of
Sciences.

\end{document}